\documentclass[12pt]{article}
\usepackage{epsfig, amssymb}
\usepackage{graphicx,epsfig}
\usepackage{color}
\usepackage{xcolor}
\begin{document}
\begin{titlepage}
\thispagestyle{empty}
\begin{flushright}
\end{flushright}

\bigskip

\begin{center}
\noindent{\Large \textbf
{First order formalism of holographic Wilsonian renormalization group: Langevin equation}}\\
\vspace{2cm} \noindent{
Jae-Hyuk Oh${}^{a}$\footnote{jaehyukoh@hanyang.ac.kr} 
}

\vspace{1cm}
  {\it
Department of Physics, Hanyang University \\
 Seoul 133-891, Korea${}^{a}$\\
 }
\end{center}

\vspace{0.3cm}
\begin{abstract}
We study a mathematical relationship between holographic Wilsonian renormalization group and stochastic quantization framework. We extend the original proposal given in arXiv:1209.2242 to interacting theories. The original proposal suggests that fictitious time(or stochastic time) evolution of stochastic 2-point correlation function will be identical to the radial evolution of the double trace operator of certain classes of holographic models, which are free theories in AdS space. We study holographic gravity models with interations in AdS space and establish a map between the holographic renormalization flow of multi-trace operators and stochastic $n$-point functions.
To give precise examples, we extensively study conformally coupled scalar theory in AdS$_6$. What we have found is that the stochastic time $t$ dependent 3-point function obtained from Langevin equation with its Euclidean action being given by $S_E=2I_{os}$ is identical to holographic renormalization group evolution of holographic triple trace operator as its energy scale $r$ changes once an identification of $t=r$ is made.
$I_{os}$ is the on-shell action of holographic model of conformally coupled scalar theory at the AdS boundary. We argue that this can be fully extended to mathematical relationship between multi point functions and multi trace operators in each framework.
\end{abstract}
\end{titlepage}

\newpage

\tableofcontents
\section{Introduction}
\label{introduction}
The best way to illustrate stochastic process is to discuss a simple example of the thermal relaxation process\cite{Parisi:1980ys,Damgaard:1987rr}
\footnote{To explain the stochastic process, we mostly follow the review, \cite{Damgaard:1987rr}}
. One of them is Brownian motion. To describe the Brownian motion, one may consider a statistical velocity profile in 1-dimension or statistical velocity vector bundle's profile in $d$-dimension($d>1$). The velocity profile satisfies Langevin equation, which is given by
\begin{equation}
m\frac{d \vec v(t)}{dt}=-\alpha \vec v(t) + \vec \eta(t),
\end{equation}
where $\vec v$ is the velocity of a test particle, which is injected into a thermal bath with its temperature $T$ and the $\alpha$ is a friction constant being positive. The $\vec\eta$ provides an interaction between the test particle and the thermal bath, which is called the stochastic force vector. The stochastic force vector $\vec \eta$ shows Gaussian distribution as
\begin{equation}
\label{GAussian-noise-FUNK}
{\rm Statistical\ distribution\ of\ \vec\eta}=\frac{\exp\left(-\frac{1}{4\lambda}\int d t \ \vec\eta(t)\cdot \vec\eta(t)\right)}{\int [\mathcal D\eta]\exp\left(-\frac{1}{4\lambda}\int d t \ \vec\eta(t)\cdot \vec\eta(t)\right)},
\end{equation}
which is called probability distribution, where the $\lambda\equiv\alpha k_B T$ and the $k_B$ is the Boltzman constant. The solution of the Langevin equation is given by
\begin{equation}
\vec v(t)=e^{-\frac{\alpha}{m}t}v(0)+\frac{1}{m}\int^t_0 e^{\frac{\alpha}{m}(t'-t)}\vec \eta(t')dt'.
\end{equation}
By using the solution one can evaluate a statistical expectation value of the kinetic energy as
\begin{equation}
\langle K \rangle=\frac{1}{2}m\langle \vec v(t) \cdot \vec v(t)\rangle=\frac{3}{2}k_BT\left(1-e^{-\frac{2\alpha}{m}t}\right)+\frac{1}{2}m\vec v(0)\cdot \vec v(0)e^{-\frac{2\alpha}{m}t}.
\end{equation}
The expectation value of the kinetic energy approaches to $\langle K \rangle=\frac{3}{2}k_BT$ regardless of its initial velocity, $v(0)$ as $t\rightarrow \infty$. This means that the injected test particle is to become in a statistical equilibrium state for large time. This is also because the stochastic process is Markovian which means that the collision between the test particle and the particles in the bath washes out the information of the test particle before the collision.

One can apply such a stochastic process to understand Euclidean field theories.
In this context, statistical correlations of a field $\phi$ is equivalent with the computations of the following path integrals:
\begin{equation}
\label{path-INTgggtral}
\langle \phi(x_1,t)...\phi(x_n,t)\rangle=\frac{\int [\mathcal D \phi]P(\phi(x,t),t) \phi(x_1,t)...\phi(x_n,t)}{\int [\mathcal D \phi]P(\phi(x,t),t)},
\end{equation}
where $P(\phi,t)$ is called probability distribution becoming weight of the field integration.
The field, $\phi$ satisfies Langevin equation,
\begin{equation}
\frac{\partial \phi(x,t)}{\partial t}=-\frac{1}{2}\frac{\delta S_E}{\delta \phi(x,t)}+\eta(x,t),
\end{equation} 
where the action, $S_E$ is a Euclidean theory of the field $\phi$ that we want to get its correlation functions.\footnote{To derive the path integral, we use an identical relation of stochastic partition function, 
\begin{equation}
\mathcal Z=\int[\mathcal D \phi]P(\phi(x,t),t)=\int[\mathcal D \eta]P(\eta(x,t),t),
\end{equation}
through the field redefinition by using Langevin equation.}
In fact,  the correlation functions of the field $\eta$ are given by
\begin{equation}
\langle \eta(x_1,t)...\eta(x_n,t)\rangle=\frac{\int [\mathcal D \eta]P(\eta(x,t),t) \eta(x_1,t)...\eta(x_n,t)}{\int [\mathcal D \eta]P(\eta(x,t),t)},
\end{equation}
where
\begin{equation}
P(\eta(x,t),t)=\exp\left(-\frac{1}{2}\int d t\int d^dx \ \eta(x,t)^2\right),
\end{equation}
which is Gaussian form, so $\eta$ is called white Gaussian noise field.
The correlation functions of the white Gaussian noise can be easily computed by performing Gaussian integral, and they are given by
\begin{eqnarray}
\label{noise-tt}
\langle{\rm odd\ number\ of\ \eta}\rangle=0, {\ \ }\langle \eta(x,t)\eta(x^\prime,t^\prime)\rangle=\delta(t-t^\prime)\delta^{(d)}(x-x^\prime), \\ \nonumber
\langle \eta(x_1,t)...\eta(x_n,t_n)\rangle=\sum_{\rm [all\ possible\ pair\ combinations]} \prod_{\rm [pairs]}\langle \eta(x_i,t_i)\eta(x_j,t_j)\rangle.
\end{eqnarray}

The probability distribution also satisfies a Schrodinger type equation as
\begin{equation}
\label{Fokker-PLANCK-eQ}
\frac{\partial \psi_s}{\partial t}=\mathcal H_{FP} \psi_s,
\end{equation}
where 
\begin{equation}
\label{Psi-wave-function}
\psi_s(\phi,t)=P(\phi,t)e^{S_E/2}, {\ \ \rm and \ \ }\mathcal H_{FP}=\frac{1}{2}\left(-\Pi(x)+\frac{1}{2}\frac{\delta S_E}{\delta \phi(x)}\right)\left(\Pi(x)+\frac{1}{2}\frac{\delta S_E}{\delta \phi(x)}\right).
\end{equation}
The $\Pi(x)$ is the canonical conjugate of the field $\phi(x)$, i.e. $\Pi(x)\equiv\frac{\delta}{\delta \phi(x)}$, where there is no the imaginary number, ``$\it i$'' in its definition since it is defined in Euclidean space. $\mathcal H_{FP}$ is called Fokker-Planck Hamiltonian. Even though one can start with an arbitrary probability distribution in the very early stochastic time $t$, for the very large time $t$, it gets reached a fixed point, where $\partial_t \psi_s=0$. One of the trivial solution in the fixed point is 
\begin{equation}
\left(\frac{\delta}{\delta \phi(x)}+\frac{1}{2}\frac{\delta S_E}{\delta \phi(x)}\right)\psi_s(\phi,t)=0,
\end{equation}
where the solution of $\psi_s=e^{-S_E/2}$ and this gives $P(\phi,t\rightarrow \infty)=e^{-S_E}$. In fact, in the very large time, the stochastic correlation functions are those of a Euclidean field theory of $S_E$, where we set $\hbar=1$. 

Namely, stochastic quantization is to get quantum correlation functions by considering the probability distribution $P(\phi,t\rightarrow\infty)$ as a Euclidean path integral weight with identification $k_BT=\hbar$. In fact, it is also discussed that the stochastic time evolution of the weight in the path integral, $P(\phi,t)$ for the Euclidean theory $S_E$ is equivalent to that we solve the corresponding Langevin equation,
\begin{equation}
\frac{\partial \phi(x,t)}{\partial t}=-\frac{1}{2}\frac{\delta S_E}{\delta \phi(x,t)}+\eta(x,t),
\end{equation}
together with the Gaussian distribution $P(\eta)$ of the stochastic force fields $\eta(x,t)$. 

Holographic Wilsonian renormalization group(RG) describes flows of deformations to dual gauge field theories as their energy scale changes by employing the holographic dual gravity models\cite{Heemskerk:2010hk,Faulkner:2010jy,deBoer:1999tgo,Akhmedov:1998vf}.
It turns out that the holographic Wilsonain RG equation can be decribed by Hamilton-Jacobi equation, 
\begin{equation}
\label{holo-hamilto-equ}
\mathcal H_{RG}\psi_H=\partial_r \psi_H,
\end{equation}
where $\mathcal H_{RG}$ is the Hamiltonian being obtained by Legendre transformation from a certain holographic gravity model Lagrangian. The variable $r$ is corresponding to an energy scale of the dual field theory. This equation describes the evolution of the wave function,
\begin{equation}
\label{S_BBBbb}
\psi_H=\exp(-S_B)
\end{equation} 
as the energy scale $r$ changes, where $S_B$ is a collection of the boundary deformations, which are those of a certain composite operators corresponding to normalizable mode of excitations in the dual gravity model. 

Recently, an interesting idea is proposed in \cite{Oh:2012bx,Jatkar:2013uga,Oh:2013tsa,Oh:2015xva,Moon:2017btx}. The evolution of the boundary deformations is identical to a complete different framework: stochastic quantization. For  holographic (free) gravity models in AdS space, the proposal suggests the three conditions,
\begin{itemize}
\item The stochastic time, ``$t$'' is identified with the variable, ``$r$'' which mediates holographic renormalization group energy scale,
\item The Euclidean action, $S_E$ is identified with the holographic on-shell action $I_{os}$ as $S_E=2I_{os}$,
\item The Fokker-Planck Hamiltonian, $\mathcal H_{FP}(t)$ has the same form with the holographic renormalization group Hamiltonian, $\mathcal H_{RG}(r)$. Namley, $\mathcal H_{RG}(r)=\mathcal H_{FP}(t)$, provided $r=t$.
\end{itemize}
where $t$ is the stochastic time and $I_{os}$ is the on-shell action computed in the gravity model at the conformal boundary in AdS space without any boundary counter terms. Once these three conditions are satisfied, the result is that {\it the stochastic two point correlation function precisely gives the evolution of holographic double trace deformation.}

The precise map is 
\begin{equation}
\label{SHW-relation}
\langle f_p(r) f_{p^\prime}(r) \rangle^{-1}_H=\langle f_p(t) f_{p^\prime}(t) \rangle^{-1}_S-\frac{1}{2}\frac{\delta^2 S_E}{\delta f_p(t) \delta f_{p^\prime}(t)},
\end{equation}
where $\langle f_p(t) f_{p^\prime}(t) \rangle_S$ is the stochastic 2-point correlation function and
\begin{equation}
\langle f_p(r) f_{p^\prime}(r) \rangle^{-1}_H=\frac{\delta^2 S_B}{\delta f_p(r) \delta f_{p^\prime}(r)}.
\end{equation}
$f_p(r)$ is the field in the gravity model and $f_p(t)$ is the stochastic field originally defined in $S_E(f_p)$. This relation is tested in many holographic models: massless scalar in AdS$_2$ and $U(1)$ gauge fields in AdS$_4$ \cite{Oh:2012bx}, conformally coupled scalar in AdS$_d$\cite{Jatkar:2013uga}, massive scalar in AdS$_d$\cite{Oh:2015xva} and massless and massive fermions in AdS$_d$\cite{Oh:2013tsa,Moon:2017btx}. They are all free theories.

In this paper, we study the features of mathematical relationship between stochastic quantization and holographic Wilsonian RG more in detail and extend the relation(\ref{SHW-relation}) to interacting theories. The main results are two fold. We illustrate these in order.

First, we clarify how the third condition, $\mathcal H_{RG}=\mathcal H_{FP}$ suggested above can be understood. It turns out that this condition needs not to address, but it is derived from the first two conditions, $r=t$ and $S_E=2I_{os}$ for certain classes of holographic models.
This condition is somewhat crucial since this ensures that the two evolutions of the wave function, $\psi_s$ and $\psi_H$ share the same form of solutions of Hamilton-Jacobi equations, (\ref{Fokker-PLANCK-eQ}) and (\ref{holo-hamilto-equ}) respectively. This tells us that holographic Wilsonian RG is nothing but the Fokker-Planck approach in stochastic framework. However, once the first two conditions are satisfied, then $\mathcal H_{RG}$ becomes precisely the same with $\mathcal H_{FP}$.

{To show this, we investigate the fixed points of holographic Wilsonian RG, namely $\partial_r \psi_H=0$ and then it means that $\mathcal H_{RG}\psi_H=0$ at the fixed point. This condition allows us to rewrite the dual gravity action in terms of $S_B$, which is evaluated at the fixed point. 
With this action, one can compute equation of motion, its solution of the action and finally compute its on-shell action. It turns out that the on-shell action is given by $I_{os}=\pm S_B$. Once we use the condition $S_E=2I_{os}=\pm2S_B$, and we replace every $S_B$ by such a $\pm\frac{S_E}{2}$ in the gravity model action, then what we have found is that the gravity model action becomes precisely the form of the Fokker-Planck action in the classical limit.
We note that the argument for the proof applies to all the kinds of free theories defined in AdS space \cite{Oh:2012bx,Jatkar:2013uga,Oh:2013tsa,Oh:2015xva,Moon:2017btx} and conformally coupled scalar with its self-interaction in AdS space is also applicable.

}

%
%
%
%
%
%
%
%
%

Second, we extend the relation to holographic gravity models with interactions. Especially, we study conformally coupled scalar theory in AdS$_6$. The reasons why we study this model is as follows. (1) There are several advance researches with this model\cite{Oh:2012bx,Jatkar:2013uga,Oh:2020zvm} to explore the relation(\ref{SHW-relation}). (2) Conformally coupled scalar theory has nice property. The theory in AdS space can be effectively defined in flat space of $\mathbb R^{d} \times \{0,\infty\}$, where $\mathbb R^d$ is $d$-dimensional Euclidean space. This is good in a sense that stochastic process is usually defined in such a space.
$\mathbb R^d$ is the space that the theory, $S_E$ is defined on and the stochastic time $t \in\{0,\infty\}$. 

What we have found is the relation between the evolutions of triple trace operator computed from the holographic gravity models and the stochastic 3-point functions. The precise map is given by
\begin{equation}
\label{SHW-relationdongdong}
\langle  f_{p_1}(r) f_{p_2}(r)  f_{p_3}(r)\rangle_H|^{r=t}=\langle f_{p_1}(t) f_{p_2}(t)  f_{p_3}(t)\rangle^c_S\prod_{i=1}^3\langle f_{p_i}(t) f_{-p_i}(t) \rangle^{-1}_S-\left.\frac{1}{2}\frac{\delta^3 S_E}{\delta f_{p_1}(t)\delta f_{p_2}(t) \delta f_{p_3}(t)}\right|^{f=0},
\end{equation}
where $\langle f_{p_1}(t) f_{p_2}(t)  f_{p_3}(t)\rangle^c_S$ is the stochastic (connected) 3-point correlation function and
\begin{equation}
\left.\langle  f_{p_1}(r) f_{p_2}(r)  f_{p_3}(r)\rangle_H=\frac{\delta^3 S_B}{\delta f_{p_1}(r)\delta f_{p_2}(r) \delta f_{p_3}(r)}\right|^{f=0}.
\end{equation}
We note that the 3-point function is evaluated up to leading order(the first order) in $\lambda$, where $\lambda$ is a coupling constant of 3-point self interaction appearing in conformally coupled scalar theory in AdS$_6$.

\section{A brief review of conformally coupled scalar in AdS space}

To discuss the conformally coupled scalar theory as a holographic model, we consider an action
\begin{equation}
\label{con-scalar-action}
S=\int_{r>\epsilon} dr d^dx \sqrt{g} \mathcal{L}(\phi ,\partial\phi)+S^\prime_B,
\end{equation}
where the theory is a probe theory in Euclidean AdS$_{d+1}$ space, whose metric is given by
\begin{equation}
\label{Euc-metric}
ds^2=g_{\mu\nu}dx^\mu dx^\nu=\frac{1}{r^2} \left( dr^2+\sum_{i=1}^{d}dx^i dx^i \right),
\end{equation}
where the indices, $\mu$, $\nu$ are $d+1$-dimensional (Euclidean)spatial indices running over 1.. to $d+1$ and the parameter $\epsilon$ is a ceratain radial cut off of AdS space. The $x^i$ are the coordinate variables along AdS boundary directions where as $r$ is the radial variable of AdS space.
The term, $S^\prime_B$ is a boundary term on the $r=\epsilon$ boundary, which will make boundary variation problem be well posed.

The detailed Lagrangian density of conformally coupled scalar theory is 
\begin{equation}
\label{Scalar-Lagrangian}
\mathcal L(\phi,\partial \phi)=\frac{1}{2}g^{\mu\nu}\partial_\mu\phi \partial_\nu\phi+\frac{1}{2}m^2 \phi^2 +\frac{\lambda}{4}\phi^\frac{2(d+1)}{d-1},
\end{equation}
For the theory to enjoy a scale symmetry, the mass $m$ needs not to be an arbitrary value, but
\begin{equation}
\label{mass-condition}
m^2=-\frac{d^2-1}{4}.
\end{equation}
In fact, the mass term comes from a coupling of the scalar field with background curvature of AdS space.
 
Such a mass plays interesting roles in holographic context.
It turns out that in a mass range of
\begin{equation}
\label{alt-condition}
-\frac{d^2}{4}\le m^2 \le -\frac{d^2}{4}+1,
\end{equation}
both of the non-normalizable and normalizable modes of the excitations in the gravity theory (in AdS space)can be sources of the deformations to the dual boundary field theory. The field theory operators coupled to either of the source terms become unitary in this mass range.

Another is again scaling property emerging in the value of the mass. By using this scaling property, we can perform a field redefinition as
\begin{equation}
\label{con-transform}
\phi(x^\mu)\equiv r^{\frac{d-1}{2}}f(x^\mu), 
\end{equation}
and then, the action(\ref{con-scalar-action}) with the Lagrangain density(\ref{Scalar-Lagrangian}) becomes
\begin{eqnarray}
\label{trans-action}
S&=&\int_{r>\epsilon}drd^dx \left( \frac{1}{2}\delta^{\mu\nu}\partial_\mu f(x) \partial_\nu f(x)+\frac{\lambda}{4}f^\frac{2(d+1)}{d-1}(x) \right) \\ \nonumber
&+&\frac{d-1}{2}\int d^dx\left. \frac{f^2(x)}{2r} \right\vert_{\epsilon}^\infty+S^\prime_B.
\end{eqnarray}
In this action, the theory is effectively defined in a half of the flat space, $\mathbb R^d \times [0,\infty)$.

We note that this theory shows self-interaction being proportional to $\sim f^{\frac{2(d+1)}{d-1}}$, where again $f$ is the field which is newly introduced by the relation(\ref{con-transform}). The exponent, $\frac{2(d+1)}{d-1}$ is fractional in general, but if $d=3$ or $d=5$, it becomes an integer. If someone considers quantum theory with this model, it is probably reasonable that one considers $d=3$, which will give $f^4$-theory or $d=5$ which will do $f^3$-theory.

%
%
%
%
%

\section{A review of holographic Wilsonian RG of conformally coupled $f^3$-theory}
\subsection{Derivation of Hamilton-Jacobi equation}
In this section, we will discuss holographic Wilsonian renormalization group equation for conformally coupled scalar theory. The most of our discussion is already appeared in \cite{Oh:2020zvm} and we faithfully follow the argument therein. In \cite{Oh:2020zvm}, the authors discuss the case that $d=5$ mostly, which means that the theory is massless scalar theory defined in 6-dimensional flat space. In this case, the interaction vertex is 3-point self interaction.
The theory is defined by the following form of the action in momentum space:
\begin{eqnarray}
 \nonumber
\label{phi3-bulk-action}
S&=&\int_{r>\epsilon} dr \left[  \frac{1}{2} \int d^5k d^5 k^\prime \delta^{(5)}(k+k^\prime)     \left( \partial_r f_k \partial_r f_{k^\prime} + k^2 f_k f_{k^\prime} \right)   +\frac{\lambda}{4(2\pi)^{5/2}}\int \prod_{i=1}^3 d^5 k_i f_{k_i}\delta^{(5)}\left(\sum_{j=1}^3 k_j\right)      \right]  \\
&+&S_B(\epsilon),
\end{eqnarray}
where to derive the action, we emply a Fourier transform as
\begin{equation}
f(r,x_i)=\frac{1}{(2\pi)^{5/2}}\int d^5k e^{-ik_ix_i}f_{k_i}(r).
\end{equation}
Therefore, $k_i$ is the momentum along the 5-dimensional boundary directions. The $S_B$ is the boundary term at $r=\epsilon$ near the conformal boundary. In fact, this contains even the boundary terms generated in the process of the field redefinition(\ref{con-transform}), and then it is 
\begin{equation}
S_B=S_B^\prime-\int d^5x\left. \frac{f^2(r,x)}{r} \right\vert^{r=\epsilon}.
\end{equation}


To derive Hamiltonian of the theory, we define our conjugate momentum of the field $f$ and its equations of motion as
\begin{equation}
\Pi_k\equiv\partial_r f_{-k}=\frac{\delta S_B}{\delta f_k}, {\ \ \ }\partial_r \Pi_k=k^2f_k + \frac{3\lambda}{4}\int \frac{d^5k^\prime}{(2\pi)^{5/2}}f_{k^\prime} f_{k-k^\prime},
\end{equation}
where the second equality of the first equation guarantees that variation of the action $S$ is well defined even on the $r=\epsilon$ boundary surface.
The theory $S$ will not depend on the cut-off scale, $\epsilon$. By applying this fact, we request $\frac{d S}{d \epsilon }=0$ and that leads a Hamilton-Jacobi type equation, which will describe the evolution of the  boundary term $S_B$ as the radial cut-off runs. Its form of the equation is given by
\begin{eqnarray}
\partial_\epsilon S_B(\epsilon)&=&-\frac{1}{2}\int d^5 k \left( \frac{\delta S_B}{\delta f_k(\epsilon)} \right)\left( \frac{\delta S_B}{\delta f_{-k}(\epsilon)}\right) +   \frac{1}{2} \int d^5k d^5 k^\prime \delta^{(5)}(k+k^\prime)      k^2 f_k f_{k^\prime}  \\ \nonumber
 &+&\frac{\lambda}{4}\int \prod_{i=1}^3 d^5 k_i f_{k_i}\delta^{(5)}\left(\sum_{j=1}^3 k_j\right)
\end{eqnarray}

Now we try to solve this equation. The form of the trial solution is designed to be an expansion in the  weak field of $f$ in momentum space. The precise form of the solution is given by
\begin{eqnarray}
\label{Ansatz-RG-holoht}
S_B(\epsilon)&=&\Lambda(\epsilon)+\int J_k(\epsilon) f_{-k}(\epsilon)d^5 k +   \left[\int\prod_{i=1}^2 d^5 k_i f_{k_i}(\epsilon)\right]D^{(2)}_{k_1k_2}(\epsilon)\delta^{(5)}\left(\sum_{j=1}^2 k_j\right)\\ \nonumber
&+&\sum_{n=1}^{\infty}\lambda^n\left[\int\prod_{i=1}^{n+2} d^5 k_i f_{k_i}(\epsilon)\right]D^{(n+2)}_{k_1,...,k_{n+2}}(\epsilon)\delta^{(5)}\left(\sum_{j=1}^{n+2} k_j\right).
\end{eqnarray}

Once we substitute the ansatz into the Hamilton-Jacobi equation, we obtain a series of terms with products of the field $f$ with certain momentum dependent coefficients in front of them. We stress that the Hamilton-Jacobi equation is an identical equation in the field $f$. Therefore, the coefficients in front of $f^n$ for an arbitrary $n$ on the both of the left and right hand sides of the equation should be the same. The equations for these coefficients are listed below:
\begin{eqnarray}
\label{floweq1}
\partial_\epsilon \Lambda(\epsilon)&=&-\frac{1}{2} \int d^5 k J_k(\epsilon)J_{-k}(\epsilon), \\
\label{floweq2}
\partial_\epsilon J_k(\epsilon)&=&-2J_k(\epsilon)D^{(2)}_{k,-k}(\epsilon), \\ \nonumber
\\ \nonumber
\label{floweq3}
\partial_\epsilon D^{(2)}_{(p,-p)}(\epsilon)&=&-\frac{1}{2}(4D^{(2)}_{(p,-p)}(\epsilon)D^{(2)}_{(-p,p)}(\epsilon)-p^2)-3\lambda\int d^5k J_{-k}(\epsilon)D^{(3)}_{(p,-p+k,-k)}(\epsilon) \\ 
\\
\label{floweq4}
\partial_\epsilon D^{(3)}_{(k_1,k_2,k_3)}(\epsilon)&=&\frac{1}{4(2\pi)^{5/2}}-2\left( \sum_{j=1}^3 D^{(2)}_{k_j,-k_j}\right)(\epsilon) D^{(3)}_{k_1,k_2,k_3}(\epsilon)
\\ \nonumber
&-&4\lambda\int d^5k J_{-k}(\epsilon)D^{(4)}_{(k_1,k_2,-k_1-k_2+k,-k)}(\epsilon)  \\ 
\label{floweqn}
\partial_\epsilon D^{(n)}_{(k_1,...,k_{n})}(\epsilon)&=&
-2\left(\sum_{j=1}^nD^{(2)}_{(k_j,-k_j)}\right)(\epsilon)D^{(n)}_{(k_1,...,,k_{n-1},-\sum_{j=1}^{n-1}k_j)}(\epsilon)\\ \nonumber
-\frac{1}{2}\sum_{n^\prime=1}^{n-3}(n^\prime+2)(n-n^\prime)&{\mathcal Per}&\left \{
D^{(n^\prime+2)}_{(k_1,...,k_{n^\prime+1},-\sum_{j=1}^{n^\prime+1}k_j)}(\epsilon)
D^{(n-n^\prime)}_{(k_{n^\prime+2},...,k_{n-1},-\sum_{j=1}^{n-1}k_j,\sum_{j=1}^{n^\prime+1}k_j)}(\epsilon)\right\} \\ \nonumber
&-&\lambda(n+1)\int d^5k J_{-k}(\epsilon)D^{(n+1)}_{(k_1,...,k_{n-1},k-\sum_{j=1}^{n-1}k_j,-k)}
\\
\nonumber
{\ \ \rm \ for\ \ }n\geq 4,
\end{eqnarray}
where $\mathcal Per\{\}$ denotes all possible permutations of momentum labels in the curly bracket.
\subsection{Solutions of Hamilton-Jacobi equation}
Looking at the above equations, $D^{(n)}$, the coefficients of $f^n$ is coupled to $D^{(m)}$, the coefficients of $f^m$ where $m\neq n$. Since they are coupled one another, it is rather hard to get their solutions. However, if one assumes that $J_k=0$, then (\ref{floweq3}) becomes an equation of one unknown, $D^{(2)}$ only. Once we assume that $J_k=0$, then the boundary cosmological constant $\Lambda$ becomes a constant meaning that it shows no $\epsilon$-dependence. The solutions of $D^{(m)}$ can be obtained from the solutions of $D^{(n)}$, where $n<m$.


The solutions of $D^{(n)}$ for $n\geq 2$ are given by
\begin{eqnarray}
\label{d-trace-sol}
D^{(2)}_{p,-p}(\epsilon)&=&\frac{1}{2}\frac{\partial_\epsilon f_p(\epsilon)}{f_p(\epsilon)}, \\
\label{t-trace-sol}
D^{(3)}_{(k_1,k_2,k_3	)}(\epsilon)&=&\frac{1}{4(2\pi)^{5/2}} \frac{\int^\epsilon \left( f_{k_1}(\epsilon^\prime)f_{k_2}(\epsilon^\prime)f_{k_3}(\epsilon^\prime)\right) d\epsilon^\prime+C^{(3)}_{k_1,k_2,k_3}}{f_{k_1}(\epsilon)f_{k_2}(\epsilon)f_{k_3}(\epsilon)}, \\
\label{n-trace-sol}
D^{(n)}_{(k_1,...,k_{n-1},-\sum_{j=1}^{n-1}k_j)}(\epsilon)&=&\frac{C^{(n)}}{\prod_{i=1}^n f_{k_i}(\epsilon)}-\frac{1}{2}\int ^\epsilon d\epsilon^\prime \left(\frac{\prod_{j=1}^n f_{k_j}(\epsilon^\prime)}{\prod_{l=1}^n f_{k_l}(\epsilon)}\right) \\ \nonumber
\times\sum_{n^\prime=1}^{n-3}(n^\prime+2)(n-n^\prime)&{\mathcal Per}&\left\{
D^{(n^\prime+2)}_{(k_1,...,k_{n^\prime+1},-\sum_{j=1}^{n^\prime+1}k_j)}(\epsilon)
D^{(n-n^\prime)}_{(k_{n^\prime+2},...,k_{n-1},-\sum_{j=1}^{n-1}k_j,\sum_{j=1}^{n^\prime+1}k_j)}(\epsilon)\right\},
\end{eqnarray}
where $C^{(3)}$ and $C^{(n)}$ are integration constants. The solution $f_p(\epsilon)$ is one of the followings:
%
%
%
\begin{equation}
\bar f_p(\epsilon)=\bar C_p \cosh[|p|(\epsilon-\bar \theta)], {\rm\ \ or\ \ }f_p(\epsilon)=C_p \sinh [|p|(\epsilon-\theta)],
\end{equation}
where $C_p$, $\bar C_p$, $\theta$ and $\bar \theta$ are arbitrary, momentum $p$ dependent constants. In fact, one solution can be obtained from another by employing an analytic continuation of the constants in the solution. For example, the second, $f_p(\epsilon)$ maps to the first solution, $\bar f_p(\epsilon)$, once we define that $\theta=\bar \theta+i\frac{\pi}{2}$ and $C_p=\bar C_p e^{i\frac{\pi}{2}}$. Therefore, we will use the second solution, $f_p(\epsilon)$ only for the later discussion.


The solution of 
$D^{(2)}_{p,-p}$ is given by
\begin{equation}
D^{(2)}_{p,-p}(\epsilon)=\frac{|p|}{2}\coth [|p|(\epsilon-\theta)].
\end{equation}
Again, we have another solution $D^{(2)}_{p,-p}(\epsilon)=\frac{|p|}{2}\tanh[|p|(\epsilon-\bar\theta)]$, which can be obtained by the analytic continuation.
This solution shows unique behavior as $\epsilon\rightarrow\infty$, which is understood as the infra-red fixed point of the coupling of double trace operator deformation to the boundary field theory where the operator is coupled to the boundary value of the field $f$ on the conformal boundary. As $\epsilon\rightarrow\infty$, $D^{(2)}_{p,-p}(\infty)=\frac{|p|}{2}$. 

The solution of $D^{(3)}_{k_1,k_2,k_3}$ is given by
\begin{eqnarray}
\label{HOLOLOLO-triple}
\lambda D^{(3)}_{k_1,k_2,k_3}(\epsilon)&=&\frac{\lambda}{4(2\pi)^{5/2}\prod_{i=1}^3\sinh[|k_i|(\epsilon-\theta)]}\left( \frac{C_p^{(3)}}{\prod_{i=1}^3C_{k_i}}
+\frac{1}{4}\frac{\cosh\left(\sum_{j=1}^3[|k_j|(\epsilon-\theta)]\right)}{\sum_{l=1}^3|k_l|} \right. \nonumber \\ \nonumber
 &-&\left.\frac{1}{4}\sum_{j=1}^3\frac{\cosh\left(\sum_{l=1}^3[|k_l|(\epsilon-\theta)]-2[|k_j|(\epsilon-\theta)]\right)}{\sum_{m=1}^3|k_m|-2|k_j|} \right),
\end{eqnarray}
which shows its fixed point
\begin{equation}
\lambda D^{(3)}_{k_1,k_2,k_3}(\infty)=\frac{\lambda}{4(2\pi)^{5/2}(\sum_{i=1}^3|k_i|)},
\end{equation}
 as $\epsilon\rightarrow\infty$.




The fixed points of $D^{(n)}$, where $n \geq 4$ as $\epsilon\rightarrow\infty$ are given by
\begin{eqnarray}
\nonumber
\lambda^{n-2}D^{(n)}_{(k_1,...,k_{n})}(\infty)&=&-\frac{\lambda^{n-2}}{2(\sum_{i=1}^{n}|k_i|)}
\sum^{n-3}_{n^\prime=1}(n^\prime+2)(n-n^\prime){\mathcal Per}\left\{D^{(n^\prime+2)}_{(k_1,...,k_{n^\prime+1},-\sum_{j=1}^{n^\prime+1}k_j)}(\infty)\right.\\ 
&\times& \left.D^{(n-n^\prime)}_{(k_{n^\prime+2},...,k_{n-1},-\sum_{j=1}^{n-1}k_j,\sum_{j=1}^{n+1}k_j)}(\infty)\right\}.
\end{eqnarray}  
For an explicit example, $D^{(4)}(\infty)$ is 
\begin{eqnarray}
\nonumber
\lambda^2 D^{(4)}_{k_1,k_2,k_3,k_4}(\infty)&=&-\frac{3\lambda^2}{2^5(2\pi)^5(\sum_{i=1}^4|k_i|)}\left(\frac{1}{(|k_1|+|k_2|+|k_1+k_2|)(|k_3|+|k_4|+|k_3+k_4|)}\right. \\ 
&+&\left.({k_1 \leftrightarrow k_3})+({k_1 \leftrightarrow k_4})\right)
\end{eqnarray}



We note that $J_k\neq0$ solutions are not much valid for further discussion. For more discussion on this issue, see the end of Section.4 in \cite{Oh:2020zvm}.

\section{Stochastic framework and Holographic Wilsonian RG}
\subsection{First order formulation for holographic theories}
Let us start this subsection with review of the previous works\cite{ Oh:2012bx,Jatkar:2013uga,Oh:2013tsa,Oh:2015xva,Moon:2017btx} on the research of the relation between stochastic quantization and holographic Wilsonian RG. In the series of these papers, it is suggested that the holographic Wilsonian renormalization group equation is nothing but a classical limit of Fokker-Planck approach once one identifies the classical action, $S_E$ 
with the boundary on-shell action in holographic context. As described in Sec.\ref{introduction}, probability distribution $P(f(t),t)$ in stochastic partition function satisfies a Schrodinger type equation given in (\ref{Fokker-PLANCK-eQ}) and (\ref{Psi-wave-function}). We note that $f(t)$ is the stochastic field and $t$ is stochastic time.  Holographic Wilsonian renormalization group equation show a form of Hamilton Jacobi equation. With an observation on the mathematical similarity between the two frameworks, it is suggested that holographic dual gravity model and its boundary theory can be described by stochastic frame. The most strong evidence obtained from the previous research is a fact that the radial evolution of holographic double trace operator in $r$ is completely captured by (stochastic) time evolution of stochastic two point function in $t$, where $r$ is AdS radial variable.

To derive this result, we need 3-conditions, where some quantities in stochastic framework match with those in holography.
The suggested conditions between the quantities in the both frameworks are the following:
\begin{itemize}
\item {\bf The First Condition}: The stochastic time, ``$t$'' is identified with the variable, ``$r$'' which mediates holographic renormalization group energy scale.  $r$ is AdS radial variable.
\item {\bf The Second Condition}: The Euclidean action, $S_E$ is identified with the holographic on-shell action $I_{os}$ as $S_E=2I_{os}$.
\item {\bf The Third Condition}: The Fokker-Planck Hamiltonian, $\mathcal H_{FP}(t)$ has the same form with the holographic renormalization group Hamiltonian, $\mathcal H_{RG}(r)$. Namely, $\mathcal H_{RG}(r)=\mathcal H_{FP}(t)$, provided $r=t$.
\end{itemize}

For every case that the authors look at\cite{ Oh:2012bx,Jatkar:2013uga,Oh:2013tsa,Oh:2015xva,Moon:2017btx}, such an identification, $S_E=2I_{os}$ recovers the dual gravity action(also Hamiltonian) 
generating the on-shell action as a Fokker-Planck Lagrangian(also Hamiltonian) density. In other wards, the Fokker-Planck Lagrangian density being given from $S_E$ has the same form with the holographic action if its boundary on-shell action is given by $I_{os}=2S_E$. This supports the third condition addressed above, $\mathcal H_{RG}(r)=\mathcal H_{FP}(t)$, 

In fact, the third condition is derived from the first two conditions, $t=r$ and $S_E=2I_{os}$.
This can be understood if one looks at fixed points of the holographic Wilsonian renormalization group equation. Consider the most general form of the gravity dual action of scalar fields as
\begin{equation}
\label{Dual-GrAvitty}
S=\int _{r>\epsilon} dr \left(\frac{1}{2}\int d^dk d^dk^\prime \delta^{(d)}(k+k^\prime)d_r f_k d_r f_{-k} + \mathcal L(f(r))\right)
\end{equation}
in momentum space,
where $\mathcal L(f(r))$ is the Lagrangian density which does not contain radial derivative terms. 
Namely it does not have $d_r f_k$. The symbol $d_r$ denotes $d_r=\frac{d}{dr}$
\footnote{This reason why we use the symbol $d_r$ rather than $\partial_r$ here is to use the equation(\ref{dd-we-want-use}). If some quantity, $Y$ depends on $r$ but not through explicite $r$-dependence, $\partial_rY=0$, but  $d_r Y\neq0$. In other sections, we use $\partial_r$ only.}
. We note that we assume that the term $\mathcal L(f(r))$ has no explicite $r$-dependence at all, but it depends on $r$ only through the field $f(r)$. Any free theory model of holographic gravity models in AdS space transform into this type of Lagrangian by employing an appropriate field redefinition together with redefinition of radial variable $r$(See \cite{Oh:2015xva}). Conformally coupled scalar theory with interactions are also this type.

For example, if we restrict the case to conformally coupled scalar theory in $d=5$ case (\ref{phi3-bulk-action}), then
\begin{equation} 
\mathcal L(f)= \int d^dk d^dk^\prime \delta^{(d)}(k+k^\prime) f_k f_{k^\prime}  +\frac{\lambda}{4(2\pi)^{5/2}}\int \prod_{i=1}^3 d^5 k_i f_{k_i}\delta^{(5)}\left(\sum_{j=1}^3 k_j\right).
\end{equation}
The Hamilton-Jacobi equation describing holographic Wilsonian renormalization group at classical level is given by
\begin{equation}
\partial_\epsilon S_B=\mathcal H_{RG},
\end{equation}
where
\begin{equation}
\label{Hrg-at-fixedpt}
\mathcal H_{RG}=\int_{r>\epsilon} dr\left[ 
\frac{1}{2}\int d^dk d^dk^\prime \delta^{(d)}(k+k^\prime) \left(\frac{\delta S_B}{\delta f_k}\right)\left(  \frac{\delta S_B}{\delta f_{k^\prime}}\right) - \mathcal L(f)
\right].
\end{equation}

Now, consider fixed points of the equation. Fixed points can be obtained by requesting $\partial_\epsilon S_B=0$ and then we demand that $\mathcal H_{RG}=0$ for the fixed points, too.
%
By using the relation $\mathcal H_{RG}=0$ at the fixed points, one can replace $\mathcal L(f)$ by the term, $\frac{1}{2}\int d^dk d^dk^\prime \delta^{(d)}(k+k^\prime) \left(\frac{\delta S_B}{\delta f_k}\right)\left(  \frac{\delta S_B}{\delta f_{k^\prime}}\right) $ in the holographic dual gravity action(\ref{Dual-GrAvitty}). Then, the action becomes
\begin{equation}
\label{Dual-GrAvitTTty}
S=\int _{r>\epsilon} dr \frac{1}{2}\int d^dk d^dk^\prime \delta^{(d)}(k+k^\prime)\left[d_r f_k d_r f_{-k} +\left(\frac{\delta S_B}{\delta f_k}\right)\left(  \frac{\delta S_B}{\delta f_{k^\prime}}\right)  \right].
\end{equation}
We note that we need to be careful to derive the above action. The relation  $\mathcal H_{RG}=0$ is hold only at the fixed points. However, the $\mathcal L(f)$ in (\ref{Dual-GrAvitty}) has radial variable $r$ dependence. Therefore, even if we get the relation, $\mathcal H_{RG}=0$ at the fixed point, in the action(\ref{Dual-GrAvitTTty}) we promote the field $f_{k}\rightarrow f_k(r)$ for it to become generic $r$ dependence. 

Now let us discuss equation of motion of the action(\ref{Dual-GrAvitTTty}). Variation of the action gives
\begin{equation}
\label{EOM-SBB}
d^2_rf_k(r)=\left(\frac{\delta S_B}{\delta f_{k}}\right)\left(  \frac{\delta^2 S_B}{\delta f^2_{-k}}\right).
\end{equation}
After mutiplying $d_r f_k$ to the both side of the equation of motion, we consider the following object:
\begin{equation}
d_rf_k(r) d^2_rf_{-k}(r)+d_rf_{-k}(r) d^2_rf_{k}(r)=\left(\frac{\delta S_B}{\delta f_{k}}\right)\left(  \frac{\delta^2 S_B}{\delta f^2_{-k}}\right)d_rf_{-k}(r)+\left(\frac{\delta S_B}{\delta f_{-k}}\right)\left(  \frac{\delta^2 S_B}{\delta f^2_{k}}\right)d_rf_k(r).
\end{equation}
We note that for the next step in this computation, we will use a fact that
\begin{equation}
\frac{d S_B(f_k)}{dr}= \left(\frac{\delta S_B(f_k)}{\delta f_{k}}\right)\frac{df_k}{dr}+\frac{\partial S_B}{\partial r}
\end{equation}
and the last term vanishes. Therefore, 
\begin{equation}
\label{dd-we-want-use}
\frac{d S_B(f_k)}{dr}= \left(\frac{\delta S_B(f_k)}{\delta f_{k}}\right)d_r f_{k}
\end{equation}
This is because the $S_B$ is not usual one but it evaluate at fixed points and assign $r$-dependence only on the field $f_k$. Thus, the $S_B$ has no explicit $r$-dependence at all.

By using that $\frac{\delta}{\delta f_k}d_r f_k=d_r$, this object becomes a total derivative form as
\begin{equation}
\frac{d}{d r}\left[ d_rf_k(r) d_rf_{-k}(r)-\left(\frac{\delta S_B}{\delta f_{k}}\right)\left(  \frac{\delta S_B}{\delta f_{-k}}\right)\right]=0.
\end{equation}
The left hand side of the above equation can be factorized and one can rewrite it as
\begin{equation}
 \left[d_rf_k(r)-\left(  \frac{\delta S_B}{\delta f_{-k}}\right)\right]\left[d_rf_{-k}(r)+\left(  \frac{\delta S_B}{\delta f_{k}}\right)\right]=C_k,
\end{equation}
where $C_k$ has no $r$ dependence, namely just a boundary momenta, $k$ dependent constant and $C_k=C_{-k}$.

%
%
%
%
%
%
%
%
%
%
%
%
%
The equation of motion(\ref{EOM-SBB}) is non-linear equation in the field $f_k$, and so it is rather hard to get its solution, since $S_B$ can be a generic function of the field $f_k$. However, in a case that the integration constant $C_k=0$, we have two first order equations. The two equations are
\begin{equation}
d_rf_k(r)\pm\left(  \frac{\delta S_B}{\delta f_{-k}}\right)=0.
\end{equation}
In fact, the solutions of holographic dual gravity theories obtained in \cite{ Oh:2012bx,Jatkar:2013uga,Oh:2013tsa,Oh:2015xva,Moon:2017btx} are all these kinds. 

Let us take these solutions and compute on-shell action of the dual gravity action in holographic setting(\ref{Dual-GrAvitTTty}). We manipulate the form of the action as
\begin{equation}
\label{EOM-SBB-final-onshell}
S=\int _{r>\epsilon} dr \frac{1}{2}\int d^dk d^dk^\prime \delta^{(d)}(k+k^\prime)\left[\left(d_r f_k  \pm\frac{\delta S_B}{\delta f_{k^\prime}}\right)\left( d_r f_{k}\pm  \frac{\delta S_B}{\delta f_{k^\prime}}\right) \mp \frac{d}{dr}(2S_B)\right].
\end{equation}
The first term in the above action vanishes thanks to the equation of motion, and the second term is boundary term at $r=\epsilon$. Therefore, up to equation of motion, we get holographic boundary on-shell action as 
\begin{equation}
I_{os}=\pm S_B(r=\epsilon).
\end{equation}
Finally, we understand that once we identify $S_E=2I_{os}=\pm2S_B$ recalling the second condition that we address in the beginning of this subsection, the action(\ref{Dual-GrAvitTTty}) becomes Fokker-Planck action in the classical limit if we set $r=t$(the first conditon). 
The form of the Fokker-Planck Lagrangian is given by
\begin{equation}
\mathcal L_{FP}=\frac{1}{2}\left(\frac{df_k(t)}{dt}\right)\left(\frac{df_{-k}(t)}{dt}\right)
+\frac{1}{8}\left(\frac{\delta S_E}{\delta f_k(t)}\right)\left(\frac{\delta S_E}{\delta f_{-k}(t)}\right)-\frac{1}{4}\left(\frac{\delta^2 S_E}{\delta f_k(t) \delta f_{-k}(t)}\right),
\end{equation}
and the last term is known as higher order in $\hbar$, where we set $\hbar=1$\cite{Parisi:1980ys,Damgaard:1987rr}. Thus, we ignore the last term in the following discussion.
This also means that the Hamilton-Jacobi equation for holographic Wilsonian renormalization group is nothing but Fokker-Planck equation(\ref{Fokker-PLANCK-eQ}).
%
%
%
%
%
%
%
%
%
%

So far, we justify that the identification $S_E=2I_{os}$ provide a feature that the holographic Wilsonian renormalization group equation is the same with the Fokker-Planck Hamiltonian equation.  According to this result, we understand that the solution of each Hamiltonian equation of wave function should be the same in the classical limit. The wave function of  holographic Wilsonian renormalization group equation $\psi_H=\exp(-S_B)$ can be identified to the wave function of the Fokker-Planck equation $\psi_S= P(f_k)\exp(\frac{S_E}{2})$, where $P(f_k)$ is the probability distribution in stochastic framework. The forms of the wave functions are given in Sec.\ref{introduction}. 

Let us develop our discussion on the probabilition distribution.
The stochastic $n$-point correlation function is given by
\begin{equation}
\left \langle\prod_{i=1}^n f_{k_i}(t)\right\rangle_s=\int [\mathcal D f_k(t)] P(f_k(t);t)\prod_{l=1}^n f_{k_l}(t),
\end{equation}
where the $f_p(t)$ is the stochastic field to be quantized. The probability distribution is defined to be
$ P(f(t),t)= e^{-S_p(f(t),t)}$, where the $S_p$ is the weight for the correlations, being given by
\begin{equation}
\label{SP-deltaoils}
S_p(f(t),t)\equiv \sum_{i=2}^\infty \left[\prod_{j=1}^i \int f_{k_j}(t)d^5k_j\right]P_i(k_1,...k_i;t)\delta^{(5)}\left( \sum_{l=1}^ik_j \right),
\end{equation}
where $P_i$ are coefficients in front of $n$-multiples of the field $f_k$.
Recalling the discussion given in \cite{Oh:2012bx} and as addressed previously, we identify the two objects: $\psi_H=\exp(-S_B)$ and $\psi_S= P(f_k)\exp(\frac{S_E}{2})$. 
In the classical limit,
\begin{equation}
S_B=S_p-\frac{1}{2}S_E.
\end{equation}
This relation is translated into a form of
\begin{equation}
\label{The-crucial relation}
\left.\lambda^{n-2}D^{(n)}_{k_1,...,k_n}(\epsilon)\right|_{\epsilon=t}=P_n(k_1,...,k_n;t)-\frac{1}{2}G_n(k_1,...,k_n;t),
\end{equation}
where $G_n$ is given by
\begin{equation}
S_E=\sum^\infty_{n=2}\int \left[ \prod_{i=1}^{n} f_{k_i}d^5k_{i} \right]{ \mathcal O_n(k_1,...,k_n)},
\end{equation}
where
\begin{equation}
\mathcal O(k_1,...,k_n)={ G_n(k_1,...,k_n)}\delta^{(d)}\left(\sum_{i=1}^n k_i\right).
\end{equation}
We also use (\ref{Ansatz-RG-holoht}) and (\ref{SP-deltaoils}) to derive this.

\subsection{Stochastic correlation functions and the precise map of the two and three point functions with double and triple trace operators in holographic Wilsonian renormalizaion group.}

In this subsection, we will develop the presice maps for the 2- and 3- point functions in stochastic framework and double and triple trace operators given in (\ref{SHW-relation})-(\ref{SHW-relationdongdong}). We start with the stochastic partition functon(\ref{path-INTgggtral}), which is given by
\begin{eqnarray}
\mathcal Z=\int [\mathcal Df_k]e^{-S_p}=\int [\mathcal Df_k]\exp\left[ -\int P_2(k_1,k_2)\delta^{(d)}(k_1+k_2)\prod_{i=1}^2f_{k_i}d^5k_i \right.\\ \nonumber
 \left.-\int P_3(k_1,k_2,k_3)\delta^{(d)}(k_1+k_2+k_3)\prod_{i=1}^3f_{k_i}d^5k_i]-...+\int J_k f_k d^5 k\right],
\end{eqnarray}
where we assume that the interaction, $P_n$ for $n>2$ is suppressed by a small coupling constant in it. Namely, we suppose that $P_n(k_i.t)\sim O(g^{n-2})$ for a certain small coupling constant $g$. The last term is a source term coupled to the field $f_k$. We expand the partition function up to first order in $P_3$, then we have
\begin{eqnarray}
\nonumber
\mathcal Z&=&\int [\mathcal Df_k]\left\{1-\int P_3(k_1,k_2,k_3)\delta^{(d)}(k_1+k_2+k_3)\prod_{i=1}^3\frac{\delta}{\delta J_{k_i}}d^5k_i]+ {\rm higher\ order\ interactions}\right\}\\ 
 &\times&\exp\left[ -\int P_2(k_1,k_2)\delta^{(d)}(k_1+k_2)\prod_{i=1}^2f_{k_i}d^5k_i+ \int J_k f_k d^5 k\right],
\end{eqnarray}
where we repalce every $f_k$ by $\frac{\delta}{\delta J_k}$ in the expansion of curly bracket. After this, we integrate the field $f_k$, then we have
\begin{eqnarray}
\nonumber
\mathcal Z&=& \left\{1-\int P_3(k_1,k_2,k_3)\delta^{(d)}(k_1+k_2+k_3)\prod_{i=1}^3\frac{\delta}{\delta J_{k_i}}d^5k_i]+ {\rm higher\ order\ interactions}\right\}\\ 
 &\times&\exp\left[  -\frac{1}{4}\int d^5 p_1 d^5 p_2\frac{\delta^{(5)}(p_1+p_2)}{P_2(p_1,p_2;t)}J_{p_1}(t)J_{p_2}(t)   \right].
\end{eqnarray}

Now, we are ready to compute correlation functions. The 2-point correlation function is
\begin{equation}
\langle f_{k_1}f_{k_2} \rangle_S=\frac{\delta^2 \log \mathcal Z}{\delta J_{k_1}\delta J_{k_2}}=\frac{1}{2P_2(k_1,k_2;t)}\delta^{(5)}\left(\sum_{i=1}^2k_i\right)
\end{equation}
and the 3-point function is
\begin{equation}
\langle f_{k_1}f_{k_2} f_{k_3} \rangle_S=\frac{\delta^3 \log \mathcal Z}{\delta J_{k_1}\delta J_{k_2}\delta J_{k_3}}=3! P_3(k_1,k_2,k_3;t)\prod_{i=1}^3\frac{1}{2P_2(k_i,-k_i;t)}\delta^{(5)}\left(\sum_{i=1}^3k_i\right).
\end{equation}
These correlation functions are obtained up to leading order in the implicit coupling constant $g$. Namely, $\langle f_{k_1}f_{k_2} \rangle_S\sim O(1)$ and $\langle f_{k_1}f_{k_2} f_{k_3} \rangle_S\sim O(g)$. We ignore all the other subleading corrections. We also ignore the tadpole diagrams which are not connected one. Inverse relations of these are
\begin{equation}
P_2(k_1,k_2;t)=\frac{1}{2}\langle f_{k_1}f_{k_2} \rangle_S, {\rm\ \ and\ \ } P_3(k_1,k_2,k_3;t)=\frac{1}{3!}\langle f_{k_1}f_{k_2} f_{k_3} \rangle_S\prod_{i=1}^3\langle f_{k_i}f_{-k_i} \rangle_S^{-1},
\end{equation}
where we drop the delta-functions.
By using a fact that
\begin{equation}
\lambda^{n-2} D^{(n)}_{k_1,...,k_n}=\frac{1}{n!}\frac{\delta^n S_B}{\delta f_{k_1}...\delta f_{k_n}}{\rm \ \ and \ \ }
G_{n}(k_1,...,k_n)=\frac{1}{n!}\frac{\delta^n S_E}{\delta f_{k_1}...\delta f_{k_n}},
\end{equation}
the relation(\ref{The-crucial relation}) for $n=2$ and $n=3$ cases are translated into
\begin{equation}
\langle f_p(r) f_{p^\prime}(r) \rangle^{-1}_H|^{r=t}=\langle f_p(t) f_{p^\prime}(t) \rangle^{-1}_S-\frac{1}{2}\frac{\delta^2 S_E}{\delta f_p(t) \delta f_{p^\prime}(t)},
\end{equation}
and
\begin{equation}
\label{SHW-relationdonghhhdong}
\langle  f_{p_1}(r) f_{p_2}(r)  f_{p_3}(r)\rangle_H|^{r=t}=\langle f_{p_1}(t) f_{p_2}(t)  f_{p_3}(t)\rangle_S\prod_{i=1}^3\langle f_{p_i}(t) f_{-p_i}(t) \rangle^{-1}_S-\left.\frac{1}{2}\frac{\delta^3 S_E}{\delta f_{p_1}(t)\delta f_{p_2}(t) \delta f_{p_3}(t)}\right|^{f=0},
\end{equation}
respectively, where
\begin{equation}
\langle f_p(r) f_{p^\prime}(r) \rangle^{-1}_H=\frac{\delta^2 S_B}{\delta f_p(r) \delta f_{p^\prime}(r)},
\end{equation}
and
\begin{equation}
\left.\langle  f_{p_1}(r) f_{p_2}(r)  f_{p_3}(r)\rangle_H=\frac{\delta^3 S_B}{\delta f_{p_1}(r)\delta f_{p_2}(r) \delta f_{p_3}(r)}\right|^{f=0}.
\end{equation}





\section{Stochastic 3-point function and check the relation(\ref{SHW-relationdonghhhdong})}

In this section, we interpret the holographic description of the renormalization group in the language of stochastic quantization. 
\subsection{Construction of Euclidean action, $S_E$}
The Fokker-Planck Lagrangian that we employ is given by
\begin{equation}
\mathcal L_{FP}=\frac{1}{2}\partial_t f_p \partial_t f_{-p} + \frac{1}{8}\left(\frac{\delta S_E}{\delta f_p}\right)\left(\frac{\delta S_E}{\delta f_{-p}}\right)-\frac{1}{4}\frac{\delta^2 S_E}{\delta f_p\delta f_{-p}},
\end{equation}
The last term is a higher order in $\hbar$, which corresponds to the quantum correction of the theory. We ignore this term for further discussion.
Now, we define the Euclidean action as
\begin{equation}
\label{Sc}
S_E=\sum^\infty_{n=2}\int \left[ \prod_{i=1}^{n} f_{k_i}d^5k_{i} \right]{\mathcal O_n(k_1,...,k_n)},
\end{equation}
where the coefficients, $O_n(k_1,...,k_n)$ are unknowns yet and will be determined soon. There are two different ways to get the exact form of 
 $O_n(k_1,...,k_n)$. One way is to get them by using the second condition that $S_E=2I_{os}$ for the corresponding stochastic system by computing the on-shell action, $I_{os}$ explicitly, which is proposed in \cite{Oh:2012bx,Jatkar:2013uga,Oh:2013tsa,Oh:2015xva,Moon:2017btx}.

The other way to obtain this is 
by employing the condition of $\mathcal H_{RG}=0$ at the fixed points in holographic RG equation where $\mathcal H_{RG}$ is given in (\ref{Hrg-at-fixedpt}) together with the fact that  the on-shell action is given by $I_{os}=\pm S_B$ at the conformal boundary. Namely,  $S_E=2I_{os}=\pm 2S_B$.
We use this method in the following and it turns out that which is completely identical with the first method.
We replace every $S_B$ by $\frac{S_E}{2}$ in the Hamiltonian $\mathcal H_{RG}$ given in (\ref{Hrg-at-fixedpt}) and request  $\mathcal H_{RG}=0$, then we have
\begin{equation}
\label{find-euclidean-action}
\frac{1}{8}\int drd^5k\left(\frac{\delta S_E}{\delta f_k}\right)\left(\frac{\delta S_E}{\delta f_{-k}}\right)=
\int drd^5k \left( \frac{1}{2}k^2 f_kf_{-k} + \frac{\lambda}{4(2\pi)^{5/2}}\int f_k f_{k^\prime} f_{-k-k^\prime} dk^\prime \right),
\end{equation}
where
\begin{eqnarray}
\int &drd^5k&\left(\frac{\delta S_E}{\delta f_k}\right)\left(\frac{\delta S_E}{\delta f_{-k}}\right)= \\ \nonumber
\int &drd^5q& \sum_{N=2}^\infty \sum_{n=2}^N \int \left[ \prod_{i=1}^{N} f_{k_i}dk_{i} \right] n(N+2-n) \mathcal Per\{\mathcal O_n(k_1,...,k_{n-1},q)\mathcal O_{N-n+2}(k_n,...,k_N,-q)\}.
\end{eqnarray}
To derive the above equation, we assume that the operator $\mathcal O_n(k_1,...,k_n)$ is permutation invariant under exchanging of the momentum labels.

To find the correct form of the Euclidean action, we request that the relation(\ref{find-euclidean-action}) is identically satisfied. Since the form of the Euclidean action is comprised of a power expansion of the field $f_p$, we request the coefficient of the $n$-multiples of the field $f_p$ i.e. the coefficient of $\left[ \prod_{i=1}^{n} f_{k_i}dk_{i} \right]$ in both sides of the relation(\ref{find-euclidean-action}) are the same. For example, the coefficients of the bilinear of the field $f_p$ in the both sides of the equation(\ref{find-euclidean-action}) are identified as
\begin{equation}
\frac{1}{2}\left[ \prod_{i=1}^{2} f_{k_i}d^5k_{i} \right]\int d^5qdr\mathcal Per \mathcal\{O_2(k_1,q)\mathcal O_2(k_2,-q)\}=
\frac{1}{2}\left[ \prod_{i=1}^{2} f_{k_i}d^5k_{i} \right]\int d^5qdr|k_1||k_2|\delta^{(5)}(k_1+q)\delta^{(5)}(k_2-q),
\end{equation}
and then,
\begin{equation}
\mathcal O_2(k_1,k_2)=\frac{|k_1|+|k_2|}{2}\delta^{(5)}(k_1+k_2)=\frac{\sum_{i=1}^2|k_i|}{2}\delta^{(5)}\left(\sum_{i=1}^2k_i\right),
\end{equation}
which is manifestly invariant under permutations of the momentum labels. Next we look at $\mathcal O_{3}(k_1,k_2,k_3)$. The tri-linear term in the left hand side of the equation(\ref{find-euclidean-action}) is given by
\begin{equation}
\frac{1}{2}\left[ \prod_{i=1}^{3} f_{k_i}d^5k_{i} \right]\int dr\left(\sum_{i=1}^3|k_i|\right)\mathcal O_3({k_1,k_2,k_3})=
\frac{\lambda}{4(2\pi)^{5/2}}\left[ \prod_{i=1}^{3} f_{k_i}d^5k_{i} \right]\int dr\delta^{(5)}\left(\sum_{i=1}^3 k_i\right).
\end{equation}
From this, we get
\begin{equation}
\mathcal O_3(k_1,k_2,k_3)=\frac{\lambda}{2(2\pi)^{5/2}(\sum_{i=1}^3|k_i|)}\delta^{(5)}(\sum_{i=1}^3k_i)
\end{equation}

The right hand side of the equation(\ref{find-euclidean-action}) does not have terms containing $\prod_{i=1}^{N} f_{k_i}d^5k_{i}$ factor, when $N \geq 4$. Therefore, the general equation to obtain the multi-linear terms, $\mathcal O_N(k_1,...,k_N)$ for $N\geq4$ is given by
\begin{equation}
0=
\int drd^5q  \sum_{n=2}^N \int \left[ \prod_{i=1}^{N} f_{k_i}dk_{i} \right]n(N+2-n) {\ }\mathcal Per \{\mathcal O_n(k_1,...,k_{n-1},q)\mathcal O_{N-n+2}(k_n,...,k_N,-q)\}.
\end{equation}
We pick out the N-multiple terms, $\mathcal O_N$ and separate it from others, then the equation becomes 
\begin{eqnarray}
\nonumber
0&=&\delta^{(5)}\left(\sum_{i=1}^Nk_i\right)\left[4\left(\sum_{i=1}^N|k_i|\right)G_N(k_1,...,k_N)+\sum_{n=3}^{N-1}n(N+2-n)\right.\\ 
&\times&\mathcal Per\{\left.G_n(k_1,...,k_{n-1},q)G_{N+2-n}(k_n,...,k_N,-q)\}\right]
\end{eqnarray}
where we define the quantity $G_n$ as
\begin{equation}
\label{Gn-define}
\mathcal O_n({k_1,...,k_n})\equiv\delta^{(5)}\left(\sum_{i=1}^nk_i\right)G_n(k_1,...,k_n).
\end{equation}
We will not compute every $\mathcal O_N$ explicitly except the $\mathcal O_4$ below.
The tetra-linear term is obtained by solving the above equation as
\begin{eqnarray}
\nonumber
G_4(k_1,k_2,k_3,k_4)&=&-\frac{3\lambda^2}{16(2\pi)^5(\sum_{i=1}^4|k_i|)}\left(\frac{1}{(|k_1|+|k_2|+|k_1+k_2|)(|k_3|+|k_4|+|k_3+k_4|)}\right. \\ 
&+&\left.({k_1 \leftrightarrow k_3})+({k_1 \leftrightarrow k_4})\right)
\end{eqnarray}



\subsection{Evaluation of stochastic 3-point correlation}
The Langevin equation is given by
\begin{equation}
\frac{\partial f_p(t)}{\partial t}=-\frac{1}{2}\frac{\delta S_E}{\delta f_{-p}(t)}+\eta_p(t),
\end{equation}
where the Euclidean action is given by (\ref{Sc}) and (\ref{Gn-define}). The $\eta_p(t)$ is the stochastic noise function satisfying (\ref{noise-tt}).
We plug the Euclidean action into the Langevin equation, we have
\begin{equation}
\frac{\partial f_p(t)}{\partial t}=-\frac{1}{2}\sum^\infty_{n=2}n \left[\int \prod_{i=1}^{n-1} f_{k_i}d^5k_{i} {\ }{\mathcal O_n(k_1,...,k_{n-1},-p)} \right]
+\eta_p(t).
\end{equation}
Since there are many of the terms on the right hand side of the equation, we solve the equation with a power expansion order by order in the small coupling $\lambda$.
The quantity, $O_{n}(k_1,...,k_n)$ is suppressed by $\lambda^{n-2}$ as seen in the solutions of them being obtained previously.
The trial solution of the equation is
\begin{equation}
f_{p}(t)=\sum_{n=0}^\infty f_p^{(n-2)}(t),
\end{equation}
where we assume that $ f_p^{(n-2)}(t)$ is an object in order of $\lambda^{n-2}$.

More practical form of the equation for the generic $n$ is given by
\begin{equation}
\label{LaNG-practical}
\frac{\partial f^{(n-2)}_p(t)}{\partial t}=-\frac{1}{2}{\sum^n_{n'=2}} n'{\sum_{n_1,n_2,...,n_{n'-1}=2}^{\infty}}'\left[ \int \prod_{i=1}^{n'-1}\left\{ f^{(n_i-2)}_{k_i}d^5k_{i}\right\}  {\mathcal O_{n'}(k_1,...,k_{n'-1},-p)} \right]+\eta_p(t),
\end{equation}
where $\sum'$(the second sum) denotes that it is a summation satisfying a condition that 
\begin{equation}
n=2-n'+\sum_{i=1}^{n'-1}n_i,
\end{equation}
and we assume that the Gaussian noise, $\eta$ is an $O(\lambda^0)$ quantity.

Now we evaluate the solutions of the equation(\ref{LaNG-practical}) step by step below.
First, we consider $n=0$ or $1$. In these cases, the form of the equation is given by
\begin{equation}
\frac{\partial f^{(0)}_p(t)}{\partial t}=\frac{\partial f^{(1)}_p(t)}{\partial t}=0
\end{equation}
and their solutions are given by
\begin{equation}
f^{(-2)}_p(t)=\Lambda(p)\lambda^{-2}{,\rm\ \ \ }f^{(-1)}_p(t)=J(p)\lambda^{-1}
\end{equation}
where $\Lambda(p)$ and $J(p)$ are arbitrary momentum dependent $O(\lambda^0)$ functions. For a moment, we assume that they vanish, i.e.$\Lambda(p)=J(p)=0$ .

Second, we consider the solution in $O(\lambda^0)$, namely $n=2$ case. In this case, the Langevin equation becomes
\begin{equation}
\frac{\partial f^{(0)}_p(t)}{\partial t}=-G_2(p,-p)f^{(0)}_{p}(t)
+\eta_p(t),
\end{equation}
and the most general solution of the equation is
\begin{equation}
\label{ffff-00-solU}
f^{(0)}_p(t)=\int^t_\tau e^{-G_2(p,-p)(t-t')}\eta_p(t')dt',
\end{equation}
where $\tau$ is a constant which is introduced to set an initial boundary condition.
We just pose here a moment and get stochastic 2-point correlation function for later use, which is given by
\begin{eqnarray}
\langle f_p(t)f_{p^\prime}(t) \rangle=\int^t_\tau\int^t_\tau dt^\prime dt^{\prime\prime}\exp\left[-G_2^p(t-t^\prime)-G_2^{p^\prime}(t-t^{\prime\prime})\right]\langle \eta_p(t)\eta_{p^\prime}(t) \rangle,
\end{eqnarray}
where $\eta_p(t)$ is the Gaussian noise function(\ref{noise-tt}), satisfying
\begin{equation}
\langle \eta_{p_1}(t_1)\eta_{p_2}(t_2)\rangle=\delta^{(d)}(p_1+p_2)\delta(t_1-t_2).
\end{equation}
By using this noise correlation function, we evaluate the $t^\prime$ and $t^{\prime\prime}$ integrations, and we get
\begin{equation}
\langle f_p(t)f_{p^\prime}(t) \rangle=-\delta^{(d)}(p+p^\prime)\frac{\sinh[G_2^{p}(\tau-t)]}{G_2^{p}\exp[-G_2^{p}(\tau-t)]}.
\end{equation}

The equation in $O(\lambda)$ is given by
\begin{equation}
\frac{\partial f^{(1)}_p(t)}{\partial t}=-G_2(p,-p)f^{(1)}_{p}(t)-\frac{3}{2} \int \left[\prod^2_{i=1}f^{(0)}_{k_i}d^5 k_i \right]
\mathcal O_3(k_1,k_2;-p),
\end{equation}
and equation in $O(\lambda^2)$ is given by
\begin{eqnarray}
\nonumber
\frac{\partial f^{(2)}_p(t)}{\partial t}&=&-G_2(p,-p)f^{(2)}_{p}(t)-\frac{3}{2} \int \left[\prod^2_{i=1}d^5 k_i \right](f^{(0)}_{k_1}f^{(1)}_{k_2}+f^{(1)}_{k_1}f^{(0)}_{k_2})
\mathcal O_3(k_1,k_2;-p) \\ \nonumber
&-&2\mathcal O_4(k_1,k_2,k_3;-p) \left[\prod^3_{i=1}f^{(0)}_{k_i}d^5 k_i \right].
\end{eqnarray}
In fact, the general form of the $n$-th order equation in $\lambda$ has a form of
\begin{equation}
\label{general-LE}
\frac{\partial f^{(n)}_p(t)}{\partial t}=-G_2(p,-p)f^{(n)}_{p}(t)+\eta^{(n)}_p,
\end{equation}
where
\begin{eqnarray}
\eta^{(0)}_p&=&\eta_p, \\
\label{need1}
\eta^{(1)}_p&=&-\frac{3}{2} \int \left[\prod^2_{i=1}f^{(0)}_{k_i}d^5 k_i \right]\mathcal O_3(k_1,k_2;-p), \\
\eta^{(2)}_p&=&-\frac{3}{2} \int \left[\prod^2_{i=1}d^5 k_i \right](f^{(0)}_{k_1}f^{(1)}_{k_2}+f^{(1)}_{k_1}f^{(0)}_{k_2})
\mathcal O_3(k_1,k_2;-p)\\ \nonumber
&-&2\mathcal O_4(k_1,k_2,k_3;-p) \left[\prod^3_{i=1}f^{(0)}_{k_i}d^5 k_i \right]
{\rm \ \ and\ so \ on.}
\end{eqnarray}
The general solution of the equation(\ref{general-LE}) is given by
\begin{equation}
\label{general-LANGevin-sol}
f^{(n)}_p(t)=\int^t_\tau e^{-G_2(p,-p)(t-t')}\eta^{(n)}_p(t')dt'.
\end{equation}


To be more precise, we show some of the explicit solutions of the above equations and the stochastic correlations. We concentrate on 3-point function. The 3-point function up to its leading order in $\lambda$ is given by
\begin{eqnarray}
\nonumber
\langle f_{p_1}(t) f_{p_2}(t) f_{p_3}(t) \rangle&=&\langle f^{(0)}_{p_1}(t) f^{(0)}_{p_2}(t) f^{(1)}_{p_3}(t) \rangle + (p_1\leftrightarrow p_3)+(p_2\leftrightarrow p_3)\\ \nonumber
&=&\exp\left(-{\sum_{i=1}^3 G_2(p_i,-p_i)t}\right)
\int^t_\tau e^{ G_2{(p_1,-p_1)}t'+G_2{(p_2,-p_2)}t''+G_2{(p_3,-p_3)}t'''}dt'dt''dt'''\\ \nonumber
&\times&\left\{\langle \eta^{(0)}_{p_1}(t')  \eta^{(0)}_{p_2}(t'')  \eta^{(1)}_{p_3}(t''')\rangle+(p_1\leftrightarrow p_3,t'\leftrightarrow t''')+(p_2\leftrightarrow p_3,t''\leftrightarrow t''')\right\}, \\ 
&\ &
\end{eqnarray}
where to derive this, we use the general form of the solution of Langevin equation (\ref{general-LANGevin-sol}), and
\begin{eqnarray}
\label{etaetaeta}
\nonumber
\langle \eta^{(0)}_{p_1}(t')  \eta^{(0)}_{p_2}(t'')  \eta^{(1)}_{p_3}(t''')\rangle&=&
-\frac{3}{2} \int \left[\prod^2_{i=1}d^5 k_i \right]\mathcal \delta^{(5)}(k_1+k_2-p_3)G_3(k_1,k_2;-p_3) \\ \nonumber
&\times&\int^{t'''}_\tau dt_1 dt_2 e^{-G_2(k_1,-k_1)(t'''-t_1)-G_2(k_2,-k_2)(t'''-t_2)} \\ 
&\times&\langle \eta_{p_1}(t')  \eta_{p_2}(t'')  \eta_{k_1}(t_1) \eta_{k_2}(t_2)\rangle.
\end{eqnarray}
 The form of $\eta_p^{(1)}$ given in (\ref{need1}), the solution of $f_p^{(0)}$ given in (\ref{ffff-00-solU}) are used to derive (\ref{etaetaeta}).
By using the fact that the correlations of white Gaussian noise $\eta(t)$,
\begin{eqnarray}
\langle \eta_{p_1}(t')  \eta_{p_2}(t'')  \eta_{k_1}(t_1) \eta_{k_2}(t_2)\rangle&=&\delta^{(5)}(p_1+p_2)\delta(t'-t'')\delta^{(5)}(k_1+k_2)\delta(t_1-t_2) \\ \nonumber
&+&\delta^{(5)}(p_1+k_1)\delta(t'-t_1)\delta^{(5)}(p_2+k_2)\delta(t''-t_2)\\ \nonumber
&+&\delta^{(5)}(p_1+k_2)\delta(t'-t_2)\delta^{(5)}(p_2+k_1)\delta(t''-t_1),
\end{eqnarray}
we get
\begin{eqnarray}
\nonumber
\langle \eta^{(0)}_{p_1}(t')  \eta^{(0)}_{p_2}(t'')  \eta^{(1)}_{p_3}(t''')\rangle&=&-\frac{3}{2} \int \left[\prod^2_{i=1}d^5 k_i \right]\mathcal \delta^{(5)}(k_1+k_2-p_3)G_3(k_1,k_2;-p_3) \\ \nonumber
&\times&\left\{ \frac{1-e^{(G_2^{k_1}+G_2^{k_2})(\tau-t''')}}{G_2^{k_1}+G_2^{k_2}} \delta^{(5)}(p_1+p_2)\delta(t'-t'')\delta^{(5)}(k_1+k_2)\right. \\ \nonumber
&+&\Theta(t'''-t')\Theta(t'''-t'')\left(e^{G_2^{k_1}(t'-t''')+G_2^{k_2}(t''-t''')}\delta^{(5)}(p_1+k_1)\delta^{(5)}(p_2+k_2) \right.\\ 
&+&\left.(p_1\leftrightarrow p_2,t' \leftrightarrow t'')) \right\},
\end{eqnarray}
where we define a new symbol, $G_2^p\equiv G_2(p,-p)=|p|$.

A Long and tedious calculations take us more compact form of the 3-point function. The 3-point correlation is comprised of two parts: one is disconnected diagram, i.e. tadpole and another is connected correlation function. The tadpole diagram(disconnected) is given by
\begin{eqnarray}
\nonumber
\langle f_{p_1}(t) f_{p_2}(t) f_{p_3}(t) \rangle_{\rm tp}&=&-\frac{3}{2}
\delta^{(5)}(p_3)\delta^{(5)}(p_1+p_2)\left( \lim_{p_3 \rightarrow 0} \frac{1-e^{G_2^{p_3}(\tau-t)}}{2G_2^{p_3}}\right)\int d^5 k_1 G_3(k_1,-k_1;p_3)\frac{1}{2G_2^{k_1}} \\ 
&\times&\left(\frac{1-e^{2G_2^{p_1}(\tau-t)}}{2G_2^{p_1}} -\frac{e^{2G_2^{k_1}(\tau-t)}-e^{2G_2^{p_1}(\tau-t)}}{2(G_2^{p_1}-G_2^{k_1})} \right),
\end{eqnarray}
and the connected 3-point function is
\begin{eqnarray}
\nonumber
\langle f_{p_1}(t) f_{p_2}(t) f_{p_3}(t) \rangle^c_{S}&=&-\frac{3\cdot 2}{2}\delta^{(5)}\left(\sum_{i=1}^3 p_i\right)\frac{G_3(-p_1,-p_2;-p_3)}{4G_2^{p_1}G_2^{p_2}}
\left\{ \frac{1-\exp\left({\sum_{i=1}^3G_2^{p_i}(\tau-t)}\right)}{\sum_{i=1}^3G_2^{p_i}}  \right. \\ \nonumber
&-& \frac{\exp\left({2(G_2^{p_1}+G_2^{p_2})(\tau-t)}\right)-\exp\left({\sum_{i=1}^3G_2^{p_i}(\tau-t)}\right)}{G_2^{p_1}+G_2^{p_2}-G_2^{p_3}} \\ \nonumber
&-& \frac{\exp\left({2G_2^{p_1}(\tau-t)}\right)-\exp\left({\sum_{i=1}^3G_2^{p_i}(\tau-t)}\right)}{G_2^{p_2}+G_2^{p_3}-G_2^{p_1}} \\ \nonumber
&-&\left.\frac{\exp\left({2G_2^{p_2}(\tau-t)}\right)-\exp\left({\sum_{i=1}^3G_2^{p_i}(\tau-t)}\right)}{G_2^{p_3}+G_2^{p_1}-G_2^{p_2}} \right\} \\
&+&(p_2\leftrightarrow p_3)+(p_1\leftrightarrow p_3) ,
\end{eqnarray}
Finally we get a compact form of the connected 3-point function, which is given by
\begin{eqnarray}
\nonumber
\label{final-ex-stoch}
\langle f_{p_1}(t) f_{p_2}(t) f_{p_3}(t) \rangle^{c}_S=-\frac{3}{4}\delta^{(5)}\left(\sum_{i=1}^3 p_i\right)\frac{G_3(-p_1,-p_2;-p_3)}{[\prod_{i=1}^3 G_2^{p_i}\exp\{-G_2^{p_i}(\tau-t)\}]}\left\{ -1+\sum_{j=1}^3\left( \frac{\sum_{l=1}^3 G_2^{p_l}}{\sum_{m=1}^3 G_2^{p_m}-2G_2^{p_j}} \right) \right. \\ \nonumber
+\left. \exp\left\{-\sum_{j=1}^3 G_2^{p_j}(\tau-t) \right\}+\sum_{m=1}^3\sinh\left\{\left(\sum_{j=1}^3 G_2^{p_j}-2G_2^{p_m}\right)(\tau-t)\right\}\right.
\\ \nonumber
+\left.\sum_{j=1}^3\left( \frac{\cosh\left[\left( \sum_{n=1}^3 G_2^{p_n}-2G_2^{p_j} \right) (\tau-t) \right]}{\sum_{m=1}^3 G_2^{p_m}-2G_2^{p_j}} \right)
 \left(\sum_{l=1}^3G_2^{p_l}\right)\right\}. \\
\end{eqnarray}
Recall the relation for 3-point functions that we compute
\begin{equation}
\label{SHjjjW-relation}
\langle  f_{p_1}(r) f_{p_2}(r)  f_{p_3}(r)\rangle_H|^{r=t}=\langle f_{p_1}(t) f_{p_2}(t)  f_{p_3}(t)\rangle^c_S\prod_{i=1}^3\langle f_{p_i}(t) f_{-p_i}(t) \rangle^{-1}_S-\left.\frac{1}{2}\frac{\delta^3 S_E}{\delta f_{p_1}(t)\delta f_{p_2}(t) \delta f_{p_3}(t)}\right|^{f=0},
\end{equation}
where $\langle f_{p_1}(t) f_{p_2}(t)  f_{p_3}(t)\rangle_S$ is the stochastic 3-point correlation function and
\begin{equation}
\left.\langle  f_{p_1}(r) f_{p_2}(r)  f_{p_3}(r)\rangle_H=\frac{\delta^3 S_B}{\delta f_{p_1}(r)\delta f_{p_2}(r) \delta f_{p_3}(r)}\right|^{f=0}.
\end{equation}
To check if this quantity matches with evolution of the holographic triple trace operator in $r$, we compute the right hand side of the relation(\ref{SHjjjW-relation}), which is given by
\begin{eqnarray}
\nonumber
\label{FINal-relation-match}
{\rm R.H.S\ of\ (\ref{SHjjjW-relation})}=-\frac{3}{4}\delta^{(5)}\left(\sum_{i=1}^3 p_i\right)\frac{G_3(-p_1,-p_2;-p_3)}{\prod_{i=1}^3\sinh[G_2^{p_i}(\tau-t)]}\left\{ -1+\sum_{j=1}^3\left( \frac{\sum_{l=1}^3 G_2^{p_l}}{\sum_{m=1}^3 G_2^{p_m}-2G_2^{p_j}} \right) \right. \\ \nonumber
+\left. \cosh\left[\sum_{j=1}^3 G_2^{p_j}(\tau-t)\right]-\sum_{j=1}^3\left( \frac{\cosh\left[\left( \sum_{n=1}^3 G_2^{p_n}-2G_2^{p_j} \right) (\tau-t) \right]}{\sum_{m=1}^3 G_2^{p_m}-2G_2^{p_j}} \right)
 \left(\sum_{l=1}^3G_2^{p_l}\right)\right\}, \\
\end{eqnarray}
where we use an identity,
\begin{equation}
\prod_{i=1}^3\sinh\left\{  G_2^{p_i}(\tau-t)\right\}=\frac{1}{4}\left[\sinh\left\{\sum_{j=1}^3 G_2^{p_j}(\tau-t)\right\}-\sum_{m=1}^3\sinh\left\{\left(\sum_{j=1}^3 G_2^{p_j}-2G_2^{p_m}\right)(\tau-t)\right\}\right].
\end{equation}
The expression(\ref{FINal-relation-match}) is completely matches with the triple trace operator expression(\ref{HOLOLOLO-triple}) when $t=r$, $\tau=\theta$ and $-1+\sum_{j=1}^3\left( \frac{\sum_{l=1}^3 G_2^{p_l}}{\sum_{m=1}^3 G_2^{p_m}-2G_2^{p_j}} \right)= \frac{4C_p^{(3)}}{\prod_{i=1}^3C_{k_i}}$.


We note that in the solution of triple trace operator(\ref{t-trace-sol}), rather than introducing the integration constant $C^{(3)}$ and indefinite integration, we use definite integration with initial condition as
\begin{equation}
D^{(3)}_{(k_1,k_2,k_3	)}(\epsilon)=\frac{1}{4(2\pi)^{5/2}} \frac{\int^\epsilon_\theta \left( f_{k_1}(\epsilon^\prime)f_{k_2}(\epsilon^\prime)f_{k_3}(\epsilon^\prime)\right) d\epsilon^\prime}{f_{k_1}(\epsilon)f_{k_2}(\epsilon)f_{k_3}(\epsilon)},
\end{equation}
then, the lower limit of the integration, $\theta$ precisely reproduce the constant, $-1+\sum_{j=1}^3\left( \frac{\sum_{l=1}^3 G_2^{p_l}}{\sum_{m=1}^3 G_2^{p_m}-2G_2^{p_j}} \right)$ in (\ref{FINal-relation-match}).


\section*{Acknowledgment}
J.H.O would like to thank his $\mathcal W.J.$ and $\mathcal Y.J.$ This work was supported by the National Research Foundation of Korea(NRF) grant funded by the Korea government(MSIP) (No.2016R1C1B1010107). This work is also partially supported by Research Institute for Natural Sciences, Hanyang University.

\end{document}